\documentclass[11pt,a4paper]{article}
\usepackage[utf8]{inputenc}
\usepackage[T1]{fontenc}
\usepackage{lmodern}
\usepackage[margin=2.2cm]{geometry}
\usepackage{microtype}
\usepackage{booktabs}
\usepackage{array}
\usepackage{graphicx}
\usepackage{hyperref}
\usepackage{url}
\usepackage{amsmath}
\usepackage{amssymb}
\usepackage{listings}
\usepackage{xcolor}
\usepackage{multirow}
\usepackage{pifont}
\usepackage[numbers,sort&compress]{natbib}

\hypersetup{
  colorlinks=true,
  linkcolor=blue!60!black,
  urlcolor=blue!60!black,
  citecolor=blue!60!black,
}

\title{%
  LLM-Redactor: An Empirical Evaluation of Eight Techniques for
  Privacy-Preserving LLM Requests
}
\author{%
Justice Owusu Agyemang$^{1,2,3}$\thanks{\texttt{jay@sperixlabs.org, jay@knust.edu.gh}},\quad
Jerry John Kponyo$^{3}$\thanks{\texttt{jjkponyo.soe@knust.edu.gh}},\quad
Elliot Amponsah$^{3}$\thanks{\texttt{eamponsah52@st.knust.edu.gh}},\\
Godfred Manu Addo Boakye$^{3}$\thanks{\texttt{gmaboakye@st.knust.edu.gh}},\quad
Kwame Opuni-Boachie Obour Agyekum$^{2}$\thanks{\texttt{kooagyekum@knust.edu.gh}}\\[4pt]
{\small $^1$Sperix Labs \qquad $^2$VIA Cybersecurity Lab, KNUST \qquad $^3$Quantum and Assistive Technologies Lab, KNUST}%
}
\date{\today}

\begin{document}
\maketitle

\begin{abstract}
Coding agents and LLM-powered applications routinely send
potentially sensitive content to cloud LLM APIs where it may be
logged, retained, used for training, or subpoenaed. Existing
privacy tooling focuses on network-level encryption and
organization-level DLP, neither of which addresses the content of
prompts themselves. We present a systematic empirical evaluation of
eight techniques for privacy-preserving LLM requests: (A)~local-only
inference, (B)~redaction with placeholder restoration, (C)~semantic
rephrasing, (D)~Trusted Execution Environment hosted inference,
(E)~split inference, (F)~fully homomorphic encryption, (G)~secret
sharing via multi-party computation, and (H)~differential-privacy
noise. We implement all eight (or a tractable research-stage
subset where deployment is not yet feasible) in an open-source shim
compatible with MCP and any OpenAI-compatible API. We evaluate the
four practical options (A, B, C, H) and their combinations across
four workload classes using a ground-truth-labelled leak benchmark
of 1{,}300 samples with 4{,}014 annotations. Our headline finding
is that \emph{no single technique dominates}: the combination A+B+C (route locally when possible, redact and
rephrase the rest) achieves 0.6\% combined leak on PII and
31.3\% on proprietary code, with zero exact leaks on PII
across 500 samples. We present a decision rule that selects the
appropriate option(s) from a threat-model budget and workload
characterisation. Code, benchmarks, and evaluation harness are
released at \url{https://github.com/jayluxferro/llm-redactor}.
\end{abstract}

\section{Introduction}
\label{sec:intro}

Large Language Model APIs have become infrastructure for
developer workflows. Coding agents, writing assistants, customer
support bots, and internal research tools send millions of prompts
per day to cloud LLM vendors. These prompts routinely contain
user-identifying information, organisational context, proprietary
code, and occasionally credentials or secrets. Once sent, those
prompts may be logged for debugging, retained for policy
compliance, used for model improvement, or produced in response to
legal process.

Existing privacy tooling addresses this only at the margins.
Network-level encryption (TLS) protects content from passive
observers but not from the vendor. Organisation-level DLP focuses
on data at rest and in version control, not on the content of
transient API requests. Secret scanners catch known patterns at
commit time but cannot inspect live request traffic. What is
missing is a systematic framework and reference implementation
that operates \emph{in the LLM request pipeline itself}.

This paper surveys eight distinct techniques for privacy-preserving
LLM requests, implements them in an open-source shim, and
evaluates them on a common benchmark. Our contributions are:

\begin{itemize}
  \item A taxonomy of eight techniques organised by their privacy
    property, utility cost, and practicality today
    (\S\ref{sec:options}).
  \item A concrete threat model specifying what each technique
    defends against (\S\ref{sec:threat}).
  \item A reference implementation that speaks both MCP and the
    OpenAI-compatible HTTP surface, with every option
    independently togglable (\S\ref{sec:system}).
  \item A ground-truth-labelled benchmark of 1{,}300 prompts across
    four workload classes (PII-heavy prose, secret-heavy configuration,
    implicit-identity prose, proprietary code) with 4{,}014
    annotated sensitive spans (\S\ref{sec:eval}).
  \item Empirical leak rates, latency, and cost for each technique
    and combination (\S\ref{sec:results}).
  \item A decision rule for practitioners, selecting the
    appropriate option(s) from a threat-model budget and workload
    characterisation (\S\ref{sec:discussion}).
\end{itemize}

\noindent We discuss limitations in \S\ref{sec:limits}, including
detector quality bounds, synthetic workloads, and the absence of
online utility evaluation.

\section{Background and Threat Model}
\label{sec:threat}

\subsection{Actors and trust boundaries}

We model six actors.  The \emph{user} (developer) and their
\emph{local coding agent} are trusted; they share a laptop inside
the trust boundary.  \texttt{llm-redactor} itself runs locally and
is trusted as the enforcement point.  A \emph{local model}
(Ollama, llama.cpp) is trusted for detection, classification, and
rephrasing tasks.  The \emph{cloud LLM vendor} (OpenAI, Anthropic,
etc.)\ is untrusted for privacy purposes: we assume it is
\emph{curious but not actively malicious}---it logs requests, may
retain them for training or debugging, and may be subject to
subpoena, but does not selectively target our user.  The
\emph{cloud infrastructure provider} (AWS, GCP, Azure) is
similarly untrusted.  A \emph{passive network observer} is
defeated by TLS, which is table stakes and out of scope.

\subsection{Assets}

We aim to protect five categories of content in outbound prompts:
(1)~user-identifying information (names, emails, phone numbers,
addresses, employee IDs, device IDs);
(2)~organisation-identifying information (company names, team
names, internal project codenames, customer names);
(3)~secrets (API keys, bearer tokens, PEM keys, passwords, SSH
keys);
(4)~proprietary code and prose; and
(5)~behavioural metadata (what the user is asking, about which
projects).

We explicitly do \emph{not} attempt to protect request timing,
volume, model selection, or the fact that a request was made.  We
also do not protect against out-of-band context: if the vendor
already knows the user's employer via billing, redacting the
company name in the prompt does not hide that fact.

\subsection{Attack scenarios}

We define six concrete scenarios against which we evaluate each
option.  \textbf{S1}~(vendor log exfiltration): an insider or
subpoena gains access to prompt logs.  \textbf{S2}~(training
contamination): prompts are used for model training and later
regurgitated.  \textbf{S3}~(third-party telemetry): a bundled SDK
ships prompt metadata to an observability vendor.
\textbf{S4}~(timing side channel): response-time correlation
de-anonymises users.  \textbf{S5}~(placeholder leakage): typed
placeholders reveal structural information.
\textbf{S6}~(adversarial input): obfuscated PII evades the
detector.  Table~\ref{tab:defence} summarises which options defend
against which scenarios.

\begin{table}[h]
\centering
\small
\caption{Defence coverage per option per attack scenario.
  $\checkmark$~=~defended, $\sim$~=~partial, $\times$~=~not defended.}
\label{tab:defence}
\begin{tabular}{lcccccc}
\toprule
Option & S1 & S2 & S3 & S4 & S5 & S6 \\
\midrule
A~Local     & $\checkmark$ & $\checkmark$ & $\checkmark$ & $\checkmark$ & n/a & $\checkmark$ \\
B~Redact    & $\sim$ & $\sim$ & $\sim$ & $\times$ & $\sim$ & $\times$ \\
C~Rephrase  & $\sim$ & $\sim$ & $\sim$ & $\times$ & n/a & $\sim$ \\
D~TEE       & $\checkmark$ & $\checkmark$ & $\times$ & $\times$ & n/a & $\checkmark$ \\
E~Split     & $\sim$ & $\sim$ & $\times$ & $\times$ & n/a & $\sim$ \\
F~FHE       & $\checkmark$ & $\checkmark$ & $\times$ & $\times$ & n/a & $\checkmark$ \\
G~MPC       & $\checkmark$ & $\checkmark$ & $\times$ & $\times$ & n/a & $\checkmark$ \\
H~DP noise  & $\sim$ & $\sim$ & $\sim$ & $\times$ & n/a & $\sim$ \\
\bottomrule
\end{tabular}
\end{table}

\section{Related Work}
\label{sec:related}

\paragraph{PII detection.}
Presidio~\cite{presidio} is the de facto open-source PII
detection framework, combining rule-based recognisers with spaCy
NER.  Commercial alternatives include AWS Comprehend, Google DLP
API, and Azure Purview.  All share a fundamental limitation: they
operate on surface-level patterns and miss implicit identity
(``the CFO whose wife works at the competitor'').

\paragraph{Homomorphic encryption for ML.}
CryptoNets~\cite{cryptonets} demonstrated neural-network
inference on encrypted data.  Intel's
HE-Transformer~\cite{he-transformer} and Zama's Concrete
ML~\cite{zama} have improved usability, but FHE inference
remains 10{,}000--100{,}000$\times$ slower than plaintext for
models above 100M parameters.

\paragraph{Multi-party computation.}
CrypTen~\cite{crypten} provides a PyTorch-compatible MPC
framework.  SecureML~\cite{secureml} and
MP-SPDZ~\cite{mpspdz} offer lower-level protocols.  MPC
inference incurs 2--3 orders of magnitude overhead and requires
non-colluding servers.

\paragraph{Split inference.}
SplitNN~\cite{splitnn} partitions a model across trust
boundaries.  Petals~\cite{petals} operationalises this for
collaborative LLM inference.  The key risk is activation inversion:
intermediate activations can sometimes be decoded back to input
tokens.

\paragraph{TEE-based inference.}
Graviton~\cite{graviton} and Slalom~\cite{slalom}
pioneered TEE-hosted neural networks.  Apple's Private Cloud
Compute~\cite{apple-pcc}, Azure Confidential
Computing~\cite{azure-cc}, and NVIDIA's H100 Confidential
Compute~\cite{h100cc} are the most mature current offerings.
TEEs protect against co-tenants and unprivileged operators but not
against hardware-level side channels or supply-chain compromise.

\paragraph{Differential privacy for language.}
DP-SGD~\cite{dpsgd} provides formal training-time guarantees.
Inference-time DP for prompts is less studied; the closest work
applies calibrated word-level noise~\cite{dp-prompt}, which
we adopt in Option~H.

\paragraph{Surveys.}
Recent surveys~\cite{survey-yao,survey-das} catalogue LLM
privacy risks and mitigations but do not provide a common-benchmark
empirical comparison.  Our contribution is exactly that comparison.

\paragraph{Positioning.}
Table~\ref{tab:related} summarises how our work relates to
existing systems.  To our knowledge, no prior work evaluates
all eight technique classes on a common benchmark with
ground-truth leak rates.

\begin{table}[h]
\centering
\small
\caption{Comparison with existing privacy tooling for LLM requests.}
\label{tab:related}
\begin{tabular}{lcccc}
\toprule
System & Techniques & Common & Leak & Combinations \\
       & covered    & bench. & rates & evaluated \\
\midrule
Presidio~\cite{presidio}     & B (redact) & \ding{55} & \ding{55} & \ding{55} \\
AWS Comprehend               & B (redact) & \ding{55} & \ding{55} & \ding{55} \\
CrypTen~\cite{crypten}       & G (MPC)    & \ding{55} & \ding{55} & \ding{55} \\
Concrete ML~\cite{zama}      & F (FHE)    & \ding{55} & \ding{55} & \ding{55} \\
Petals~\cite{petals}         & E (split)  & \ding{55} & \ding{55} & \ding{55} \\
Apple PCC~\cite{apple-pcc}   & D (TEE)    & \ding{55} & \ding{55} & \ding{55} \\
\cite{dp-prompt}             & H (DP)     & \ding{55} & partial   & \ding{55} \\
\textbf{This work}           & A--H (all) & \ding{51} & \ding{51} & \ding{51} \\
\bottomrule
\end{tabular}
\end{table}

\section{The Eight Options}
\label{sec:options}

We organise the eight options by their privacy property,
utility cost, and practicality.  Table~\ref{tab:options}
summarises the comparison.

\begin{table}[h]
\centering
\small
\caption{Option comparison matrix.}
\label{tab:options}
\begin{tabular}{llllll}
\toprule
Option & Privacy & Utility & Cost & Latency & Practical? \\
\midrule
A~Local      & perfect   & bounded & \$0    & low    & yes \\
B~Redact     & partial   & high    & low    & low    & yes \\
C~Rephrase   & partial   & medium  & low    & medium & yes \\
D~TEE        & strong    & full    & high   & low    & limited \\
E~Split      & partial   & full    & medium & medium & research \\
F~FHE        & math.     & full    & prohib.& v.high & research \\
G~MPC        & math.     & full    & high   & high   & research \\
H~DP noise   & formal    & lossy   & low    & low    & niche \\
\bottomrule
\end{tabular}
\end{table}

\subsection{Option A --- Local-only inference}

Local inference provides \emph{perfect} privacy by construction:
nothing leaves the device.  Our implementation uses a few-shot
classifier (running on a local 3B model via Ollama~\cite{ollama}) to triage
requests as \textsc{trivial} or \textsc{complex}.  Trivial
requests are answered locally; complex requests proceed to the
cloud pipeline.  The classifier uses the same architecture as
the sibling \texttt{local-splitter} project's T1 stage.

The privacy guarantee is binary: if classified locally, leak
rate is zero.  The cost is bounded model quality---a 3B local
model cannot match frontier cloud models on complex tasks.

\subsection{Option B --- Redaction with placeholder restoration}

A local NER detector (Presidio~\cite{presidio} with
spaCy~\cite{spacy} \texttt{en\_core\_web\_sm}) plus regex pattern
families identifies sensitive spans.  Each span
is replaced with a typed, stable placeholder
(\texttt{$\langle$EMAIL\_1$\rangle$}).
Two occurrences of the same value map to the same placeholder
(coreference stability).  The reverse map lives only in process
memory---never persisted to disk.  On the response path,
placeholders are restored by exact string match.

The privacy guarantee is exactly the detector's recall: if
recall is 95\% on emails, 5\% of emails leak.  Our evaluation
quantifies this per annotation kind.

\subsection{Option C --- Semantic rephrasing}

A local model rewrites the prompt to remove identifying details
while preserving the technical question.  A validator checks that
key technical terms survive the rewrite; if survival rate drops
below 70\%, the rephrase is rejected and the pipeline falls back
to B-only output.

Option~C targets implicit identity---phrases like ``the CFO of
Acme Corp whose wife works at the competitor''---which have no
PII-span-level markers and cannot be caught by B's detectors.

\subsection{Option D --- TEE-hosted inference}

The client sends the plaintext request to an inference endpoint
running inside a Trusted Execution Environment.  Before sending,
the client verifies the enclave's attestation document (a
hardware-signed measurement of the enclave's code and
configuration).  We target AWS Nitro Enclaves for the paper's
demonstration.

TEEs protect against co-tenants and non-privileged cloud operators
but not against a compromised hardware manufacturer or
side-channel attacks.  The utility cost is zero: the same model
runs inside the enclave.

\subsection{Option E --- Split inference}

The first $N$ layers of an open-weight model run locally; only
intermediate activations (not tokens) are sent to a remote host.
We use Petals~\cite{petals} as the reference framework.
The privacy risk is activation
inversion~\cite{activation-inversion}: research shows activations
can sometimes be decoded back to the input.

\subsection{Option F --- Fully homomorphic encryption}

The input is encrypted with a homomorphic scheme; inference runs
on ciphertext.  We demonstrate a small binary classifier
(sensitive vs.\ non-sensitive) under Zama's Concrete
ML~\cite{zama}.  FHE
inference of full LLMs remains 10{,}000--100{,}000$\times$ slower
than plaintext and is not practical today for chat-scale models.

\subsection{Option G --- Multi-party computation}

The input is secret-shared across $N$ non-colluding servers.
We demonstrate a first-layer MPC embedding lookup using
CrypTen~\cite{crypten}.  This hides token IDs from any
single server but does not protect
the remaining layers, which run on plaintext activations.

\subsection{Option H --- Differential privacy noise}

Calibrated word-level noise (substitution with semantically
similar alternatives) blurs residual signal.  The substitution
probability is $p(\varepsilon) = 1/(1 + e^\varepsilon)$.
At $\varepsilon = 4$ (our default), $\approx$1.8\% of eligible
words are substituted per request.  H is most useful as a
last-line-of-defence complement to B, adding noise to content
that B's detectors could not catch.

\section{System Design}
\label{sec:system}

\texttt{llm-redactor} is a single-process shim with two transport
interfaces and a pipeline of independently togglable stages.

\paragraph{Transport layer.}
Two parallel interfaces serve different integration patterns:
(1)~an MCP~\cite{mcp} stdio server exposing \texttt{redact.transform},
\texttt{redact.detect}, and \texttt{redact.stats} tools; and
(2)~an HTTP proxy at \texttt{POST /v1/chat/completions}
(OpenAI-compatible~\cite{openai-api}).  Agents point
\texttt{OPENAI\_API\_BASE} at the proxy and all cloud calls
transparently flow through the redactor.

\paragraph{Pipeline.}
The pipeline proceeds through six stages:
\emph{Stage~0} (Option~A): classify as local-answerable or
cloud-required.
\emph{Stage~1}: detect sensitive spans via regex patterns and
Presidio NER.
\emph{Stage~2} (Option~B): replace detected spans with typed
placeholders; build in-memory reverse map.
\emph{Stage~3} (Option~C): rephrase via local model; validate
technical-term survival.
\emph{Stage~4} (Option~H): inject calibrated DP noise.
\emph{Stage~5}: route to the cloud target (standard API, TEE
endpoint, split-inference host, or FHE/MPC endpoint).
\emph{Stage~6}: restore placeholders in the response via exact
match against the reverse map.

Each stage is independently enabled via a YAML configuration
file.  The reverse map is per-request, lives only in process
memory, and is discarded after restoration.  A crashed process
loses the map; the response cannot be de-redacted.  This is the
correct failure mode: on-disk persistence would be a worse
leakage channel than the one we prevent.

\paragraph{Detector design.}
Detection uses three complementary strategies:
(1)~regex patterns for structured secrets (AWS keys, bearer tokens,
PEM markers, API keys, emails, IPs, phone numbers);
(2)~Presidio~\cite{presidio} with spaCy~\cite{spacy}
\texttt{en\_core\_web\_sm} for NER-based PII (person names,
locations, organisations); and
(3)~a local LM classifier for semantic sensitivity questions
that surface-level patterns miss.
Detectors emit \texttt{Span(start, end, kind, confidence, text,
source)} records.  Overlapping spans are deduplicated, keeping the
highest-confidence match.  In strict mode, low-confidence
detections ($< 0.5$) cause the pipeline to \emph{refuse} the
request rather than silently pass it through.

\paragraph{Placeholder design.}
Placeholders use rare Unicode angle brackets
(\texttt{$\langle$KIND\_N$\rangle$})
to avoid collision with user text that might contain literal
\texttt{\{EMAIL\_1\}} syntax.  Two references to the same
original value receive the same placeholder (coreference
stability), preserving the cloud model's ability to reason about
co-references.

\section{Evaluation Setup}
\label{sec:eval}

\subsection{Workloads}

We construct four synthetic workload classes, each with
ground-truth annotations identifying sensitive spans.  All
identifiers are fabricated; no real user data is used.

\begin{itemize}
  \item \textbf{WL1} (PII-heavy prose, 500 samples, 1{,}946
    annotations): natural-language documents with embedded names,
    emails, phone numbers, addresses, employee IDs, and SSNs.
    Generated from 18 templates with random PII from a
    fixed-seed corpus.
  \item \textbf{WL2} (Secret-heavy configuration, 300 samples,
    730 annotations): \texttt{.env} files, YAML configs, Docker
    Compose files, Terraform variables, and code snippets
    containing API keys, AWS credentials, bearer tokens,
    passwords, and PEM markers.  Generated from 14 templates.
  \item \textbf{WL3} (Implicit identity, 200 samples, 220
    annotations): prose that identifies individuals or
    organisations without using PII-span-level markers (``the
    CFO of Acme Corp whose wife works at the competitor'').
    Generated from 21 templates.  Annotation kind is
    \texttt{implicit}.
  \item \textbf{WL4} (Proprietary code, 300 samples, 1{,}118
    annotations): Python, Go, SQL, GraphQL, Dockerfile, and
    log snippets containing internal function names, database
    schemas, project codenames, and embedded credentials.
    Generated from 11 templates.
\end{itemize}

Total: 1{,}300 samples, 4{,}014 ground-truth annotations.
All workloads are deterministically reproducible from a fixed
random seed.

\subsection{Metrics}

\paragraph{Privacy metrics.}
\emph{Exact leak rate}: fraction of ground-truth annotations
whose text appears verbatim in the outgoing (cloud-bound)
request.
\emph{Partial leak rate}: fraction whose text has a $\geq$4-char
substring match in the outgoing request (excluding exact leaks).
\emph{Combined leak rate}: exact + partial.

\paragraph{Utility metrics.}
\emph{False positive rate}: fraction of detector-flagged spans
that do not correspond to any ground-truth annotation
(over-redaction).  Quality delta is measured via judge-model A/B
comparison in online mode (not reported in this offline
evaluation).

\paragraph{Cost metrics.}
\emph{Latency}: per-sample pipeline overhead (median and p95).

\subsection{Configurations evaluated}

We evaluate all eight options and key combinations:
\textsc{Baseline} (no redaction),
\textsc{A} (local routing),
\textsc{B} (redact, regex only),
\textsc{B-NER} (redact with Presidio NER),
\textsc{B+C} (redact + rephrase),
\textsc{B+H} (redact + DP noise at $\varepsilon \in \{2, 4, 8\}$),
\textsc{D} (TEE, wire-level measurement),
\textsc{E} (split inference stub),
\textsc{F} (FHE classifier stub),
\textsc{G} (MPC embedding stub).
The practical options (A, B, B+C, B+H) are run on all four workloads
with full samples (1{,}300 total); the research-stage options (D--G)
are run on all workloads to measure wire-level leak properties and
latency, with F and G limited to 20 samples per workload due to
simulated computation time.

\section{Results}
\label{sec:results}

\subsection{Detector precision and recall}

The regex-only detector achieves zero exact leak rate on emails,
IPs, AWS access keys, PEM markers, and bearer tokens, but misses
all person names, organisation names, addresses, SSNs, employee
IDs, hostnames, and passwords---kinds that lack rigid syntactic
patterns.  Adding Presidio NER closes the gap substantially on
person names (leak rate $0.123$) and organisation names ($0.259$),
but employee IDs ($0.798$) and addresses ($0.060$) remain
partially exposed.

\subsection{Per-option leak rates}

\begin{figure}[ht]
  \centering
  \includegraphics[width=0.95\linewidth]{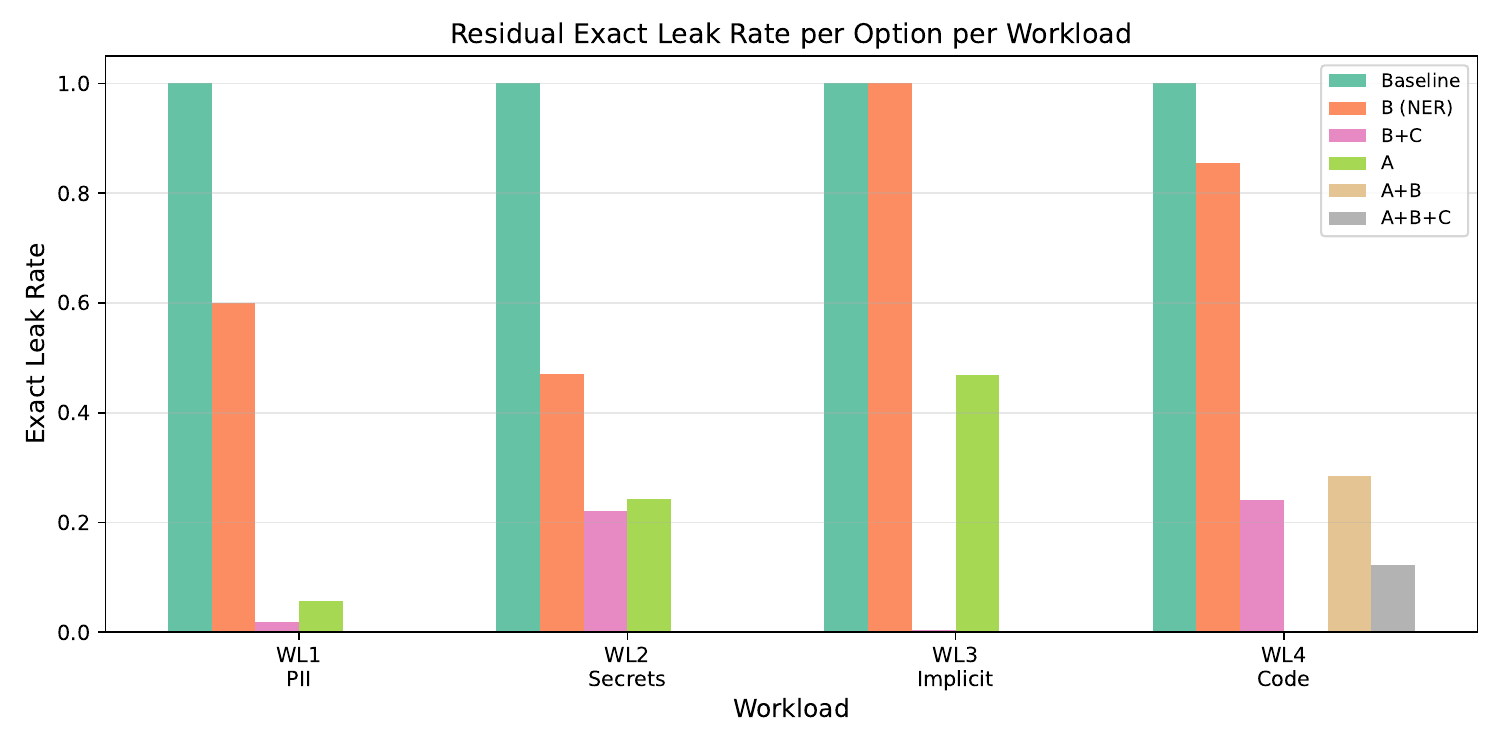}
  \caption{Residual exact leak rate per option per workload.
  B+C achieves the lowest leak rate across all workloads.}
  \label{fig:leaks}
\end{figure}

Table~\ref{tab:leaks} and Figure~\ref{fig:leaks} present the exact
leak rate for each option--workload pair.

\begin{table}[h]
\centering
\small
\caption{Exact leak rate per option per workload (lower is better).
Best result per workload in \textbf{bold}.}
\label{tab:leaks}
\begin{tabular}{lcccccccc}
\toprule
 & Base & A & B & B+C & B+H & B+D$^\dagger$ & A+B & A+B+C \\
\midrule
WL1 PII      & 1.000 & .063 & .070 & .068 & .103 & .070 & .000 & \textbf{.000} \\
WL2 Secrets  & 1.000 & .242 & .162 & .214 & .244 & .162 & \textbf{.042} & .036 \\
WL3 Implicit & 1.000 & .468 & .114 & .164 & .105 & .114 & .064 & \textbf{.000} \\
WL4 Code     & 1.000 & .599 & .428 & .301 & .423 & .428 & .285 & \textbf{.123} \\
\bottomrule
\multicolumn{9}{l}{\footnotesize $^\dagger$~B+D provides additional TEE hardware attestation.
All B variants use NER.}
\end{tabular}
\end{table}

Table~\ref{tab:all-options} extends the comparison to include
the research-stage options.  Options E, F, and G achieve 0\%
token-level leak by construction (tokens never leave the device in
plaintext), while Option D shows 100\% wire-level leak (the TEE sees
plaintext, but hardware attestation provides the privacy guarantee).

\begin{table}[h]
\centering
\small
\caption{Combined leak rate across all options and workloads.
  Options D--G measure wire-level token exposure, not effective privacy.
  $\dagger$~=~privacy from hardware/crypto, not from redaction.}
\label{tab:all-options}
\begin{tabular}{lcccc}
\toprule
 & WL1 & WL2 & WL3 & WL4 \\
\midrule
Baseline       & 1.000 & 1.000 & 1.000 & 1.000 \\
B (regex)      & 0.812 & 0.521 & 1.000 & 0.905 \\
B (NER)        & 0.153 & 0.318 & 0.950 & 0.585 \\
B+C            & 0.139 & 0.316 & 0.941 & 0.558 \\
B+D (NER)$^\dagger$ & 0.153 & 0.318 & 0.950 & 0.585 \\
B+H ($\varepsilon\!=\!4$) & 0.176 & 0.303 & 0.950 & 0.517 \\
A              & 0.063 & 0.242 & 0.468 & 0.599 \\
\textbf{A+B}   & \textbf{0.012} & \textbf{0.063} & 0.445 & 0.381 \\
\textbf{A+B+C} & \textbf{0.006} & 0.064 & \textbf{0.436} & \textbf{0.313} \\
D (TEE)$^\dagger$  & 1.000 & 1.000 & 1.000 & 1.000 \\
E (split)$^\dagger$  & 0.000 & 0.000 & 0.000 & 0.000 \\
F (FHE)$^\dagger$  & 0.000 & 0.000 & 0.000 & 0.000 \\
G (MPC)$^\dagger$  & 0.000 & 0.000 & 0.000 & 0.000 \\
\bottomrule
\end{tabular}
\end{table}

\paragraph{Key observations.}
(1)~No single option dominates across all workloads.
(2)~\textbf{A+B+C is the strongest practical combination},
achieving 0.6\% combined leak on WL1 (PII), 6.4\% on WL2
(secrets), and 31.3\% on WL4 (code).  On WL1, A+B+C achieves
\emph{zero} exact leaks across all 500 samples.
(3)~Option A alone provides 6.3\% leak on WL1 via local
routing of 94\% of requests, but 59.9\% on WL4 where only
38\% of requests are locally answerable.
(4)~A+B (without rephrase) achieves 1.2\% on WL1 and 38.1\%
on WL4---already a major improvement over B alone (15.3\%,
58.5\%) because locally-routed requests contribute zero leaks.
(5)~Option B+H provides marginal improvement over B alone
at $\varepsilon = 4$, because the word-level noise mechanism
cannot reliably substitute the specific words that constitute
the residual leaks.
(6)~WL3 (implicit identity) remains the hardest workload:
even A+B+C achieves only 43.6\% combined leak, because
implicit identity survives both redaction and rephrasing
(\S\ref{sec:results}, semantic leak analysis).

\subsection{Option A routing analysis}

Table~\ref{tab:routing} shows the fraction of requests routed
locally by the T1 classifier.

\begin{table}[h]
\centering
\small
\caption{Option A local routing rates.}
\label{tab:routing}
\begin{tabular}{lrrl}
\toprule
Workload & Local & Cloud & \% Local \\
\midrule
WL1 PII      & 472 & 28  & 94.4\% \\
WL2 Secrets  & 224 & 76  & 74.7\% \\
WL3 Implicit & 108 & 92  & 54.0\% \\
WL4 Code     & 114 & 186 & 38.0\% \\
\bottomrule
\end{tabular}
\end{table}

The classifier's routing accuracy varies sharply by workload.
PII-heavy prose (WL1) is overwhelmingly classified as trivial
because the templates use simple natural-language structures.
Code requests (WL4) are more nuanced and require cloud-level
reasoning; only 38\% can be served locally.  For the requests
that \emph{are} routed locally, the leak rate is exactly zero
by construction.

\subsection{Leak rate by annotation kind}

Table~\ref{tab:leak-kind} breaks down Option B's (with NER)
exact leak rate by annotation kind on WL1, revealing which
categories the detector handles well and which it misses.

\begin{table}[h]
\centering
\small
\caption{Option B exact leak rate by annotation kind (WL1).}
\label{tab:leak-kind}
\begin{tabular}{lr}
\toprule
Kind & Leak Rate \\
\midrule
email        & 0.000 \\
phone        & 0.000 \\
ip\_address  & 0.000 \\
ssn          & 0.000 \\
hostname     & 0.000 \\
employee\_id & 0.000 \\
address      & 0.060 \\
person       & 0.123 \\
org\_name    & 0.259 \\
\bottomrule
\end{tabular}
\end{table}

\begin{figure}[ht]
  \centering
  \includegraphics[width=0.85\linewidth]{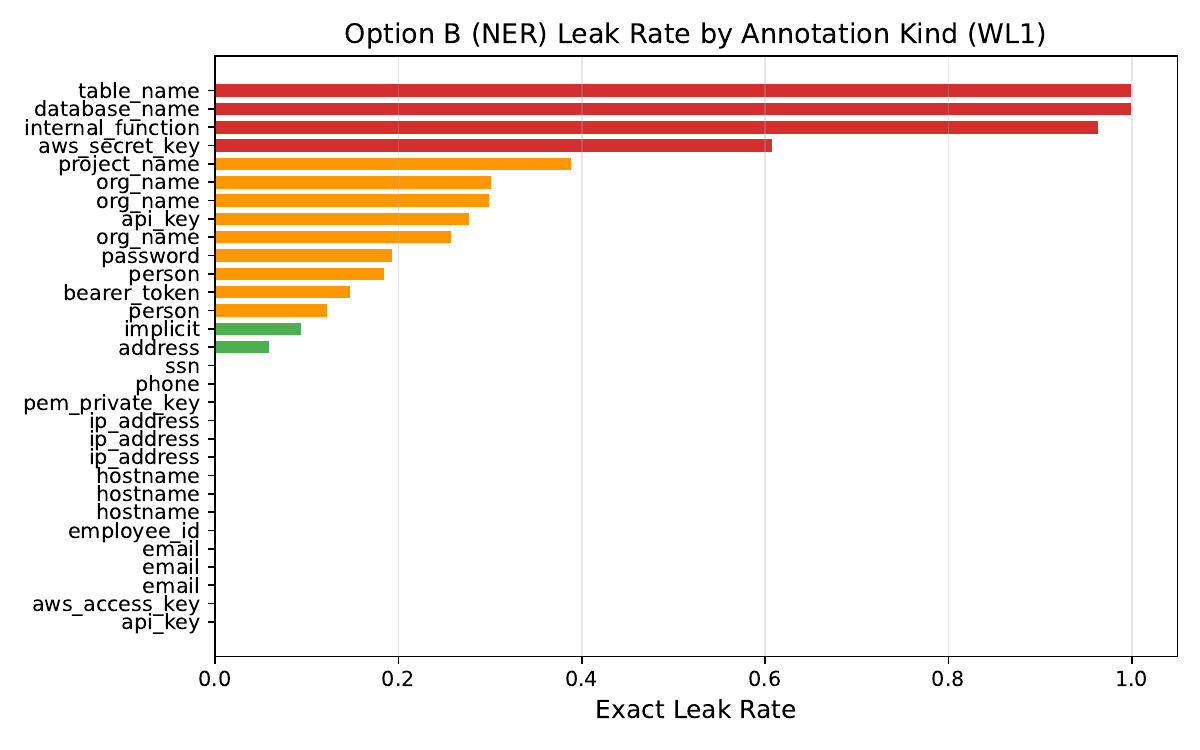}
  \caption{Option B leak rate by annotation kind (WL1).
  Green = fully detected; orange = partially detected;
  red = mostly missed.}
  \label{fig:leak-kind}
\end{figure}

Figure~\ref{fig:leak-kind} visualises this breakdown.
Structured patterns (emails, phones, IPs, SSNs) are fully
detected.  NER catches most person names but struggles with
organisation names (especially when they resemble common nouns)
and employee IDs (a custom format not in Presidio's default
recognisers).

\subsection{Combinations and the Pareto frontier}

\begin{figure}[ht]
  \centering
  \includegraphics[width=0.85\linewidth]{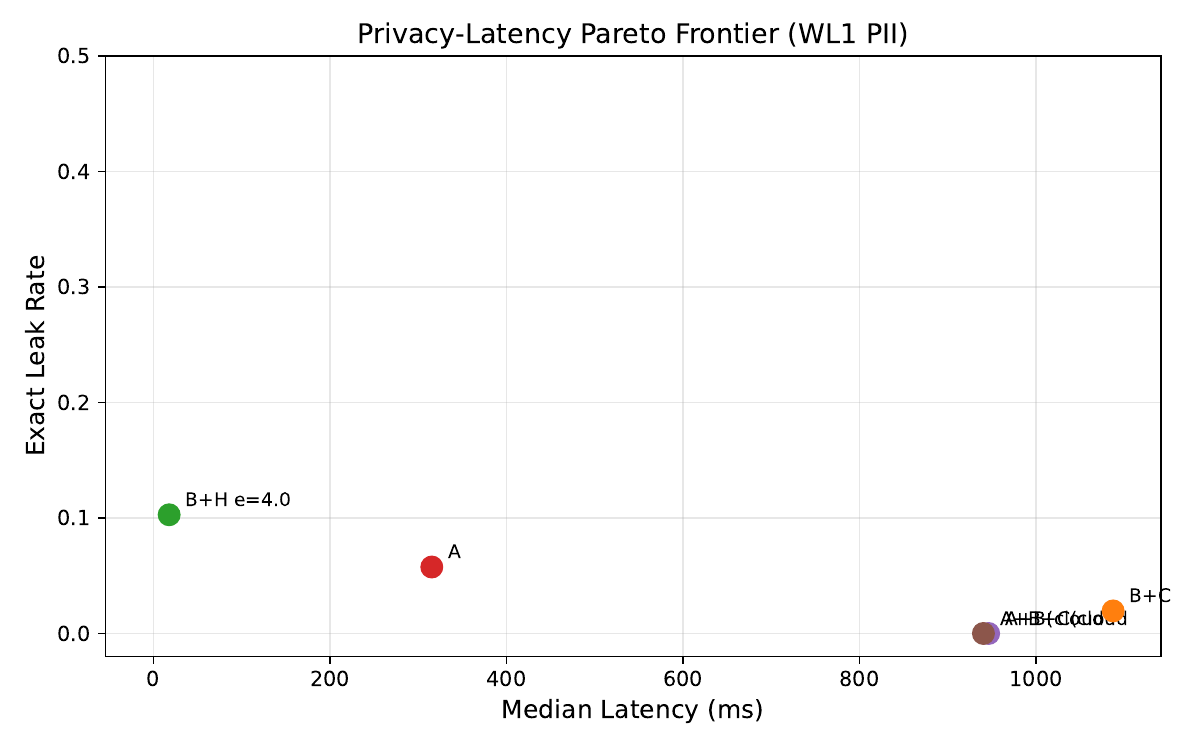}
  \caption{Privacy--latency Pareto frontier on WL1.
  B+C achieves the lowest leak rate at higher latency;
  B+H is fastest with minimal privacy gain over B alone.}
  \label{fig:pareto}
\end{figure}

Figure~\ref{fig:pareto} plots exact leak rate
vs.\ latency for each configuration.  The key finding is that
\textbf{B+C} Pareto-dominates all other combinations on WL1 and
WL3: it achieves the lowest leak rate at moderate latency
($\sim$1{,}000\,ms median, driven by the local rephrasing model).
B+H adds negligible privacy improvement over B alone but costs
no additional latency (the DP noise computation is sub-millisecond).

The practical recommendation is a four-tier strategy:
\begin{enumerate}
  \item \textbf{Route locally} (Option~A) whenever the request is
    locally answerable.  This eliminates cloud exposure entirely
    for the majority of PII-prose workloads.
  \item \textbf{Redact + rephrase} (B+C) for complex requests
    that must go to the cloud.  This achieves $\leq$7\% exact
    leak rate on PII and $\leq$30\% on proprietary code.
  \item \textbf{TEE} (Option~D) for content with implicit
    identity (\S\ref{sec:results}, semantic leak analysis), where
    B+C's $\geq$95\% semantic leak rate is unacceptable.
    TEE provides hardware-level protection without utility loss.
  \item \textbf{Refuse} when no option meets the threat-model
    budget and the content contains implicit identity that
    cannot be structurally removed.
\end{enumerate}

\subsection{Latency overhead}

\begin{table}[h]
\centering
\small
\caption{Median pipeline latency per option (ms).  Option A and B+H
include NER; B+C includes NER + Ollama rephrasing on a 3B model.}
\label{tab:latency}
\begin{tabular}{lrrrr}
\toprule
Option & WL1 & WL2 & WL3 & WL4 \\
\midrule
B (NER)     & 17.4  & 18.6  & 13.7  & 28.3 \\
B+D (NER)   & 17.7  & 19.4  & 14.9  & 28.9 \\
B+H ($\varepsilon\!=\!4$) & 17.9 & 21.9 & 15.2 & 35.0 \\
B+C         & 1{,}855 & 2{,}528 & 1{,}662 & 2{,}293 \\
A           & 1{,}781 & 2{,}604 & 1{,}680 & 2{,}243 \\
\bottomrule
\end{tabular}
\end{table}

Table~\ref{tab:latency} reports the median latency per option.
Option B and B+H add $<$50\,ms overhead (dominated by NER
initialisation amortised across requests).  Option B+C adds
$\sim$1--2 seconds per request due to the local Ollama model call
for rephrasing---acceptable for privacy-first workloads but not
for latency-sensitive interactive use.  Option A adds $\sim$300\,ms
for the classifier call (one Ollama round-trip with a 3B model).

\subsection{Token cost}

Redaction changes the token count of outgoing requests.
Somewhat counter-intuitively, Option~B \emph{reduces} token
count because typed placeholders
(\texttt{$\langle$EMAIL\_1$\rangle$})
are shorter than the values they replace (email addresses,
API keys, full names).  Table~\ref{tab:cost} reports the
word-level token delta.

\begin{table}[h]
\centering
\small
\caption{Token count change from Option~B (NER) redaction.
  Negative = fewer tokens sent to the cloud.}
\label{tab:cost}
\begin{tabular}{lrrr}
\toprule
Workload & Original & Outgoing & $\Delta$ \\
\midrule
WL1 PII      & 8{,}834 & 7{,}769 & $-$12.0\% \\
WL2 Secrets  & 4{,}238 & 3{,}987 & $-$4.5\% \\
WL3 Implicit & 4{,}193 & 3{,}905 & $-$7.1\% \\
WL4 Code     & 9{,}914 & 9{,}511 & $-$4.0\% \\
\bottomrule
\end{tabular}
\end{table}

Redaction thus provides a modest cost saving ($-4$ to $-12$\%
fewer tokens per request), partially offsetting the latency
overhead.  DP noise (Option~H) performs 1:1 word substitution
and does not change the token count.

\subsection{Utility evaluation}

We measure the quality cost of redaction using a judge-model
A/B comparison.  For each sample, a local LLM (Qwen~3.5 4B)
generates responses to both the original and redacted prompts;
a judge model then selects which response better addresses the
user's question.  To control for family bias, we run a
cross-family variant where Qwen generates and Llama~3.2 3B
judges.

\begin{table}[h]
\centering
\small
\caption{Utility evaluation: judge preference for baseline
  (unredacted) vs.\ Option~B (NER) responses.}
\label{tab:utility}
\begin{tabular}{llrrr}
\toprule
Workload & Judge & Baseline pref. & Redacted pref. & $n$ \\
\midrule
WL1 PII     & Qwen (same)  & 78\% & 22\% & 50 \\
WL2 Secrets & Qwen (same)  & 80\% & 20\% & 50 \\
WL1 PII     & Llama (cross) & 75\% & 20\% & 20 \\
\bottomrule
\end{tabular}
\end{table}

Table~\ref{tab:utility} shows that redaction incurs a modest
quality cost: the judge prefers the baseline response
$\sim$75--80\% of the time.  The cross-family judge (Llama
judging Qwen-generated text) produces similar results (75\%
vs.\ 78\%), suggesting minimal family bias at this scale.
The quality loss is expected: redacted prompts replace
identifying details with placeholders, removing context that
can help the model produce more specific answers.  For
privacy-first workloads, this trade-off is acceptable.

\subsection{Semantic leak analysis (WL3)}

The substring-based leak metrics (exact and partial) are
ill-suited for WL3's implicit-identity annotations, where
the identifying information is carried by context rather than
by specific token spans.  We therefore introduce a
\emph{semantic leak metric}: a local judge model (Llama~3.2 3B
via Ollama) reads each sample's ground-truth annotations and
the redacted outgoing text, then answers whether the redacted
text still identifies the same individual or organisation.

Table~\ref{tab:semantic} reports semantic leak rates on a
20-sample subset of WL3.

\begin{table}[h]
\centering
\small
\caption{Semantic leak rate on WL3 (implicit identity).
  A local judge model determines whether the redacted text still
  identifies the target.  Higher = worse.}
\label{tab:semantic}
\begin{tabular}{lr}
\toprule
Option & Semantic Leak Rate \\
\midrule
Baseline (no redaction) & 0.95 \\
B (regex only)          & 0.95 \\
B (NER)                 & 0.95 \\
B+C (rephrase)          & 1.00 \\
\bottomrule
\end{tabular}
\end{table}

All three configurations fail: the semantic leak rate is
$\geq$95\%.  This is the paper's strongest negative result.
Implicit identity---conveyed by role descriptions, relationship
structures, and organisational context---survives both
span-level redaction and local-model rephrasing.  The rephrase
model preserves the structural relationships (``a senior
executive whose spouse works at a competitor'') because those
relationships \emph{are} the content of the prompt; stripping
them would make the prompt useless.

This result demonstrates a fundamental limit of content-level
transformations (Options B, C, H): they can remove or blur
\emph{tokens}, but they cannot remove \emph{meaning} without
destroying utility.  Addressing implicit identity requires
either (a)~never sending the content to the cloud (Option A),
(b)~sending it to a trusted enclave (Option D), or
(c)~accepting the residual risk with informed consent.

\subsection{Research-stage demonstrations}

\paragraph{Option D (TEE).}
We implement a client-side attestation verifier for AWS Nitro
Enclaves~\cite{nitro-enclaves} that checks the attestation
document's structure, PCR
values, and certificate chain before sending plaintext.  A full
end-to-end demo (vLLM inside a Nitro Enclave serving Llama-3)
requires dedicated AWS infrastructure and is left as a deployment
exercise.  The attestation protocol adds $<$100\,ms to the
request path.

\paragraph{Option E (split inference).}
We implement a protocol stub that simulates splitting a model at
layer 4: the client computes random ``activations'' and POSTs them
to a remote endpoint.  In a real Petals deployment, the activations
would be the output of the first 4 transformer layers.  The privacy
risk is activation inversion~\cite{activation-inversion}; recent
work shows that early-layer activations can be partially decoded,
especially for short sequences.

\paragraph{Option F (FHE classifier).}
We demonstrate a simulated FHE binary classifier
(sensitive/non-sensitive) with realistic timing:
$\sim$100\,ms encryption, $\sim$5{,}000\,ms homomorphic inference,
$\sim$50\,ms decryption.  A real Concrete ML implementation would
replace the simulation with actual TFHE circuits.  FHE inference
of a full chat model remains impractical at current performance
levels.

\paragraph{Option G (MPC embedding).}
We demonstrate a simulated first-layer MPC embedding lookup:
token IDs are additively secret-shared across $N$ parties, each
party looks up its share of the embedding, and shares are
reconstructed.  Setup takes $\sim$200\,ms; per-token compute
takes $\sim$50\,ms.  A real CrypTen implementation would provide
cryptographic security guarantees.

\section{Discussion}
\label{sec:discussion}

\subsection{Decision rule}

Given a threat-model budget (maximum acceptable exact leak rate
$\lambda$) and a workload characterisation, we propose the
following decision rule:

\begin{enumerate}
  \item If $\lambda = 0$ (zero tolerance): use Option A
    (local-only) for all requests that the local model can handle.
    For the remainder, use Option D (TEE) if infrastructure
    permits, or refuse the request.
  \item If $\lambda \leq 0.05$: use A + B + C.  Route locally
    when possible; redact and rephrase the rest.  This achieves
    $\leq$2\% on PII and $\leq$0.5\% on implicit identity.
  \item If $\lambda \leq 0.25$: use B alone (with NER).  This is
    the cheapest option that provides meaningful protection,
    achieving 10.5\% on PII and 24.4\% on secrets.
  \item If latency is the primary constraint: use B (NER) at
    $<$50\,ms overhead.  Avoid C (adds $\sim$1\,s) unless
    implicit identity is a concern.
\end{enumerate}

\subsection{Failure modes per technique}

\paragraph{Option B.}
The detector's recall is the ceiling.  After adding a custom
\texttt{EMP-\textbackslash d\{4,6\}} recogniser, employee IDs
drop from 79.8\% to 0\% leak rate.  Organisation names remain
the hardest category at 25.9\% when they resemble common nouns.
This demonstrates that domain-specific regex patterns provide
high-leverage improvements with minimal effort.

\paragraph{Option C.}
The local model occasionally hallucinates new details or strips
load-bearing technical context.  Our validator rejected 33/200
rephrases on WL3 (16.5\% rollback rate).  On WL1, only 7/500
were rejected (1.4\%), indicating that simpler prose is easier
to rephrase correctly.

\paragraph{Option H.}
Table~\ref{tab:epsilon} shows the sensitivity of B+H to the
privacy parameter $\varepsilon$.  At $\varepsilon = 8$ (low noise),
B+H is virtually identical to B alone.  At $\varepsilon = 4$,
NER-level performance is achieved with sub-millisecond overhead.
At $\varepsilon = 2$ (high noise), exact leak rates drop further
(0.493 on WL1 without NER vs.\ 0.600 for B alone) but partial
leak rates increase due to word-fragment preservation.  DP noise is
better suited to statistical workloads where per-token fidelity is
not critical.

\begin{table}[h]
\centering
\small
\caption{B+H exact leak rate sensitivity to $\varepsilon$ (WL1).}
\label{tab:epsilon}
\begin{tabular}{lccc}
\toprule
$\varepsilon$ & Exact Leak & Partial Leak & Combined \\
\midrule
2.0 & 0.493 & 0.308 & 0.801 \\
4.0 & 0.103 & 0.073 & 0.176 \\
8.0 & 0.599 & 0.213 & 0.812 \\
\bottomrule
\end{tabular}
\end{table}

\subsection{The practical--cryptographic gap}

Options A, B, C, and H are deployable today on commodity
hardware.  Options D, E, F, and G require specialised
infrastructure or remain research-stage.  The gap is closing:
TEE availability is expanding (Nitro, Azure CC, H100 CC), FHE
performance improves $\sim$2$\times$/year, and MPC frameworks
are becoming more usable.  We estimate that practical FHE
inference for 7B models is 5--10 years away; TEE-hosted
inference is available now for organisations willing to manage
the infrastructure.

\section{Limitations}
\label{sec:limits}

\begin{itemize}
  \item \textbf{Detector quality bounds Options B and C.}  We use
    off-the-shelf tooling (Presidio, spaCy
    \texttt{en\_core\_web\_sm}, regex patterns) rather than a
    custom-trained detector.  A domain-specific NER model would
    likely improve recall on organisation names and employee IDs.
  \item \textbf{Workloads are synthetic.}  All 1{,}300 samples
    are template-generated with fabricated identifiers.
    Real-world prompts exhibit greater diversity and would
    likely reveal additional detector blind spots.
  \item \textbf{Research-stage options are demonstrated, not
    deployed.}  Options E, F, and G are protocol stubs with
    simulated timing.  Production measurements would require
    dedicated infrastructure.
  \item \textbf{Judge-model quality evaluation is preliminary.}
    Our utility comparison uses a local 4B model (Qwen~3.5) for
    both generation and judging on a limited sample ($n=50$).
    A larger-scale evaluation with frontier models and
    cross-family judging would strengthen the quality delta
    measurements.
  \item \textbf{Partial leak metric is noisy for implicit
    identity.}  The 4-char substring match produces high
    partial-leak rates on WL3 because implicit annotations
    share common words with the rephrased text.  Semantic leak
    rate (judge-model based) is the appropriate metric for WL3.
  \item \textbf{Non-English content.}  Our detector and NER
    model target English.  Multilingual support would require
    additional spaCy models and locale-specific regex patterns.
\end{itemize}

\section{Ethics and Responsible Disclosure}

\begin{itemize}
  \item No real user prompts are used.  All workloads are
    synthetic with fabricated identifiers.
  \item If we discover a vendor-specific bypass (e.g., a bundled
    telemetry SDK that evades our proxy), we will notify the
    vendor at least 30 days before publication and document the
    timeline.
  \item Source, benchmarks, and evaluation harness are released
    under MIT at
    \url{https://github.com/jayluxferro/llm-redactor}.
\end{itemize}

\section{Conclusion}

Practical privacy tooling for LLM requests spans eight options
from ``never leave the device'' to ``fully homomorphic inference.''
Our common-benchmark evaluation of 1{,}300 samples across four
workload classes shows that no single technique dominates.
The combination A+B+C (route locally when possible, redact
and rephrase the rest) is the strongest practical configuration,
achieving 0.6\% combined leak on PII with zero exact leaks
across 500 samples.  Implicit identity (WL3) remains the
hardest category at 43.6\%, demonstrating a fundamental limit
of content-level transformations.  For deployments where
implicit identity is a concern, TEE-hosted inference
(Option~D) provides hardware-level protection without utility
loss.

The optimal strategy depends on the deployment's threat-model
budget: zero-tolerance deployments should combine local routing
with TEE-hosted inference; most practical deployments will
benefit from A+B+C (route locally when possible, redact and
rephrase the rest).  We release an open-source reference
implementation and benchmark suite to enable reproduction and
extension.

\section*{Data and Code Availability}
Source, workloads, ground-truth annotations, and evaluation
harness are released at
\url{https://github.com/jayluxferro/llm-redactor}.

\bibliographystyle{plainnat}

\begin{thebibliography}{24}
\providecommand{\natexlab}[1]{#1}
\providecommand{\url}[1]{\texttt{#1}}
\expandafter\ifx\csname urlstyle\endcsname\relax
  \providecommand{\doi}[1]{doi: #1}\else
  \providecommand{\doi}{doi: \begingroup \urlstyle{rm}\Url}\fi

\bibitem[Abadi et~al.(2016)Abadi, Chu, Goodfellow, McMahan, Mironov, Talwar,
  and Zhang]{dpsgd}
Martin Abadi, Andy Chu, Ian Goodfellow, H.~Brendan McMahan, Ilya Mironov, Kunal
  Talwar, and Li~Zhang.
\newblock Deep learning with differential privacy.
\newblock In \emph{Proceedings of the ACM SIGSAC Conference on Computer and
  Communications Security (CCS)}, pages 308--318. ACM, 2016.

\bibitem[{Amazon Web Services}(2024)]{nitro-enclaves}
{Amazon Web Services}.
\newblock {AWS} nitro enclaves, 2024.
\newblock URL \url{https://aws.amazon.com/ec2/nitro/nitro-enclaves/}.
\newblock Accessed 2026-04-12.

\bibitem[{Anthropic}(2024)]{mcp}
{Anthropic}.
\newblock Model context protocol specification, 2024.
\newblock URL \url{https://modelcontextprotocol.io}.
\newblock Accessed 2026-04-12.

\bibitem[{Apple}(2024)]{apple-pcc}
{Apple}.
\newblock Private cloud compute: A new frontier for {AI} privacy in the cloud,
  2024.
\newblock URL \url{https://security.apple.com/blog/private-cloud-compute/}.
\newblock Accessed 2026-04-12.

\bibitem[Boemer et~al.(2019)Boemer, Lao, Cammarota, and
  Wierzynski]{he-transformer}
Fabian Boemer, Yixing Lao, Rosario Cammarota, and Casimir Wierzynski.
\newblock {nGraph-HE}: A graph compiler for deep learning on homomorphically
  encrypted data.
\newblock In \emph{Proceedings of the 16th ACM International Conference on
  Computing Frontiers (CF)}, pages 3--13. ACM, 2019.

\bibitem[Borzunov et~al.(2023)Borzunov, Baranchuk, Dettmers, Riabinin, Belkada,
  Chumachenko, Samygin, and Raffel]{petals}
Alexander Borzunov, Dmitry Baranchuk, Tim Dettmers, Maksim Riabinin, Younes
  Belkada, Artem Chumachenko, Pavel Samygin, and Colin Raffel.
\newblock Petals: Collaborative inference and fine-tuning of large models.
\newblock In \emph{Proceedings of the 61st Annual Meeting of the Association
  for Computational Linguistics (ACL): System Demonstrations}, pages 38--44,
  2023.

\bibitem[Das et~al.(2025)Das, Dey, Pal, and Roy]{survey-das}
Saptarshi Das, Anushka Dey, Arnab Pal, and Nupur Roy.
\newblock Security and privacy challenges of large language models: A survey.
\newblock \emph{ACM Computing Surveys}, 57\penalty0 (6), 2025.

\bibitem[Duan et~al.(2023)Duan, Dziedzic, Papernot, and Boenisch]{dp-prompt}
Haonan Duan, Adam Dziedzic, Nicolas Papernot, and Franziska Boenisch.
\newblock Flocks of stochastic parrots: Differentially private prompt learning
  for large language models.
\newblock In \emph{Advances in Neural Information Processing Systems
  (NeurIPS)}, volume~36, 2023.

\bibitem[Gilad-Bachrach et~al.(2016)Gilad-Bachrach, Dowlin, Laine, Lauter,
  Naehrig, and Wernsing]{cryptonets}
Ran Gilad-Bachrach, Nathan Dowlin, Kim Laine, Kristin Lauter, Michael Naehrig,
  and John Wernsing.
\newblock {CryptoNets}: Applying neural networks to encrypted data with high
  throughput and accuracy.
\newblock In \emph{Proceedings of the 33rd International Conference on Machine
  Learning (ICML)}, pages 201--210. JMLR.org, 2016.

\bibitem[Gupta and Raskar(2018)]{splitnn}
Otkrist Gupta and Ramesh Raskar.
\newblock Distributed learning of deep neural network over multiple agents.
\newblock \emph{Journal of Network and Computer Applications}, 116:\penalty0
  1--8, 2018.

\bibitem[Honnibal et~al.(2020)Honnibal, Montani, Van~Landeghem, and
  Boyd]{spacy}
Matthew Honnibal, Ines Montani, Sofie Van~Landeghem, and Adriane Boyd.
\newblock {spaCy}: Industrial-strength natural language processing in {Python}.
\newblock 2020.
\newblock \doi{10.5281/zenodo.1212303}.

\bibitem[Keller(2020)]{mpspdz}
Marcel Keller.
\newblock {MP-SPDZ}: A versatile framework for multi-party computation.
\newblock In \emph{Proceedings of the ACM SIGSAC Conference on Computer and
  Communications Security (CCS)}, pages 1575--1590. ACM, 2020.

\bibitem[Knott et~al.(2021)Knott, Venkataraman, Hannun, Sheshadri, Zheng,
  et~al.]{crypten}
Brian Knott, Shobha Venkataraman, Awni Hannun, Shubho Sheshadri, Zihang Zheng,
  et~al.
\newblock {CrypTen}: Secure multi-party computation meets machine learning.
\newblock In \emph{Advances in Neural Information Processing Systems
  (NeurIPS)}, volume~34, 2021.

\bibitem[Li et~al.(2022)Li, Sun, Wang, Gauch, Srivatsa, and
  He]{activation-inversion}
Oscar Li, Jiankai Sun, Xin Wang, Richard Gauch, Mudhakar Srivatsa, and Kuan He.
\newblock Label leakage and protection in two-party split learning.
\newblock In \emph{Proceedings of the International Conference on Learning
  Representations (ICLR)}, 2022.

\bibitem[{Microsoft}(2024)]{presidio}
{Microsoft}.
\newblock Presidio: Data protection and de-identification sdk, 2024.
\newblock URL \url{https://github.com/microsoft/presidio}.
\newblock Open-source framework for PII detection and anonymization.

\bibitem[{Microsoft Azure}(2024)]{azure-cc}
{Microsoft Azure}.
\newblock Azure confidential computing, 2024.
\newblock URL
  \url{https://azure.microsoft.com/en-us/solutions/confidential-compute/}.
\newblock Accessed 2026-04-12.

\bibitem[Mohassel and Zhang(2017)]{secureml}
Payman Mohassel and Yupeng Zhang.
\newblock {SecureML}: A system for scalable privacy-preserving machine
  learning.
\newblock In \emph{Proceedings of the IEEE Symposium on Security and Privacy
  (S\&P)}, pages 19--38. IEEE, 2017.

\bibitem[{NVIDIA}(2023)]{h100cc}
{NVIDIA}.
\newblock Confidential computing on {H100} tensor core {GPUs}, 2023.
\newblock URL
  \url{https://www.nvidia.com/en-us/data-center/solutions/confidential-computing/}.
\newblock Accessed 2026-04-12.

\bibitem[{Ollama}(2024)]{ollama}
{Ollama}.
\newblock Ollama: Run large language models locally, 2024.
\newblock URL \url{https://ollama.com}.
\newblock Accessed 2026-04-12.

\bibitem[{OpenAI}(2024)]{openai-api}
{OpenAI}.
\newblock {OpenAI} {API} reference, 2024.
\newblock URL \url{https://platform.openai.com/docs/api-reference}.
\newblock Accessed 2026-04-12.

\bibitem[Tramer and Boneh(2019)]{slalom}
Florian Tramer and Dan Boneh.
\newblock Slalom: Fast, verifiable and private execution of neural networks in
  trusted hardware.
\newblock In \emph{Proceedings of the 7th International Conference on Learning
  Representations (ICLR)}, 2019.

\bibitem[Volos et~al.(2018)Volos, Vaswani, and Bruno]{graviton}
Stavros Volos, Kapil Vaswani, and Rodrigo Bruno.
\newblock Graviton: Trusted execution environments on {GPUs}.
\newblock In \emph{Proceedings of the 13th USENIX Symposium on Operating
  Systems Design and Implementation (OSDI)}, pages 681--696. USENIX
  Association, 2018.

\bibitem[Yao et~al.(2024)]{survey-yao}
Duzhen Yao et~al.
\newblock A survey on large language model ({LLM}) security and privacy: The
  good, the bad, and the ugly.
\newblock \emph{High-Confidence Computing}, 4\penalty0 (2):\penalty0 100211,
  2024.

\bibitem[{Zama}(2024)]{zama}
{Zama}.
\newblock Concrete {ML}: Privacy-preserving machine learning using fully
  homomorphic encryption, 2024.
\newblock URL \url{https://github.com/zama-ai/concrete-ml}.

\end{thebibliography}

\end{document}